\documentclass[aps,prl,twocolumn,groupedaddress,amssymb,showpacs,epsfig]{revtex4}
\usepackage{graphicx}
\usepackage{bm}
\usepackage{dcolumn}
\usepackage{amsmath}
\usepackage{amssymb}
\usepackage{epsfig}

\def \lleq {\lower0.9ex\hbox{ $\buildrel < \over \sim$} ~}
\def \ggeq {\lower0.9ex\hbox{ $\buildrel > \over \sim$} ~}

\def \om    {\Omega}

\def \om {\Omega_{0m}}

\def \beq  {\begin{equation}}
\def \eeq  {\end{equation}}
\def \ber  {\begin{eqnarray}}
\def \eer  {\end{eqnarray}}

\def\prl{{Phys.\@ Rev.\@ Lett.\ }}

\begin{document}
\newcommand{\newc}{\newcommand}

\newc{\be}{\begin{equation}}
\newc{\ee}{\end{equation}}
\newc{\ba}{\begin{eqnarray}}
\newc{\ea}{\end{eqnarray}}
\newc{\bea}{\begin{eqnarray*}}
\newc{\eea}{\end{eqnarray*}}
\newc{\D}{\partial}
\newc{\ie}{{\it i.e.} }
\newc{\eg}{{\it e.g.} }
\newc{\etc}{{\it etc.} }
\newc{\etal}{{\it et al.}}
\newc{\lcdm }{$\Lambda$CDM }
\newcommand{\nn}{\nonumber}
\newc{\ra}{\rightarrow}
\newc{\lra}{\leftrightarrow}
\newc{\lsim}{\buildrel{<}\over{\sim}}
\newc{\gsim}{\buildrel{>}\over{\sim}}

\title{Is cosmic acceleration slowing down?}
\author{Arman Shafieloo$^a$, Varun Sahni$^b$ and Alexei A. Starobinsky$^{c,d}$}
\affiliation{$^a$ Department of Physics, University of Oxford, 1 Keble Road,
Oxford, OX1 3NP, UK\\
$^b$ Inter University Centre for Astronomy and Astrophysics, Post Bag 4,
Ganeshkhind, Pune, 411007, India\\
$^c$ Landau Institute for Theoretical Physics, Moscow 119334, Russia \\
$^d$ RESCEU, Graduate School of Science, The University of Tokyo,
Tokyo 113-0033, Japan}
\date{\today}

\begin{abstract}
We investigate the course of cosmic expansion in its {\em recent past} using the
Constitution SN Ia sample (which includes CfA data at low redshifts), jointly with
signatures of baryon acoustic oscillations (BAO) in the galaxy distribution
 and fluctuations in the cosmic microwave background (CMB).
Earlier SN Ia data sets could not address this issue because of a paucity
of data at low redshifts.
Allowing the equation of state of dark energy (DE) to vary, we find that a coasting model of
the universe ($q_0=0$) fits the data about as well as $\Lambda$CDM.
This effect, which is most clearly seen using the recently
introduced {\it Om} diagnostic, corresponds to an increase of
$Om$ and $q$ at redshifts $z\lleq 0.3$. In geometrical terms, this suggests
that cosmic acceleration may have already peaked and that we are currently witnessing
its slowing down.
The case for evolving DE strengthens if a subsample of the Constitution set consisting of
SNLS+ESSENCE+CfA SN Ia data is analysed in combination with BAO+CMB using the same
statistical methods.
The effect we observe
could correspond to DE decaying into dark matter
(or something else). A toy
model which mimics this process agrees well with the combined SN Ia+BAO+CMB data.

\end{abstract}
\maketitle


The existence of cosmic acceleration at
redshifts less than $\sim 0.5$ appears to be well established by several
independent data sets including: SN Ia luminosity distances, cosmic
microwave background temperature anisotropy and polarization maps, and
baryon acoustic oscillations in the galaxy power spectrum.
Most recent analysis performed using the data mentioned above \cite{WMAP5},
as well as data from Chandra \cite{Xray}
and SDSS \cite{cluster} cluster catalogues, show that if the DE equation
of state (EOS) $w\equiv p_{DE}/\rho_{DE}$ is assumed to be a constant, then
there remains little room for departure of DE from the cosmological constant, since $|1+w|<0.06$ at
the $1\sigma$ confidence level.
However, in the absence of compelling theoretical models with an unevolving EOS,
one must reexamine the data impartially by removing this prior,
if one is to look for serious alternatives to the cosmological constant \cite{DE_review}.

A large sample of nearby SN Ia with $z<0.08$ has recently been published \cite{CfA}.
Adding this to the Union sample \cite{Union}, leads to the
Constitution set \cite{Constitution09} which is the largest SN Ia sample to date.
Consequently one might expect significantly better limits on
$1+w$ to be placed from this new sample.
As shown in \cite{Constitution09} the Constitution set
with a BAO prior
 leads to approximately the same upper limit on $|1+w|$ as that obtained from
all other data sets if $w=const$ is assumed. However, as noted earlier,
the assumption of a constant EOS is not very realistic. For this reason, in this letter,
we drop this assumption when analyzing data from the Constitution set, together
 with BAO data at
$z=0.2$ and $z=0.35$ \cite{BAO09} and the observed CMB
shift parameter $R$ obtained from acoustic oscillations in the
CMB temperature anisotropy power spectrum \cite{WMAP5}.
The data is analyzed using the popular CPL
ansatz
\cite{CPL}
together with the recently introduced $Om(z)$ diagnostic \cite{Om,zunkel_clarkson}:

\begin{figure}
\begin{center}
\vspace{-0.1cm}
\psfig{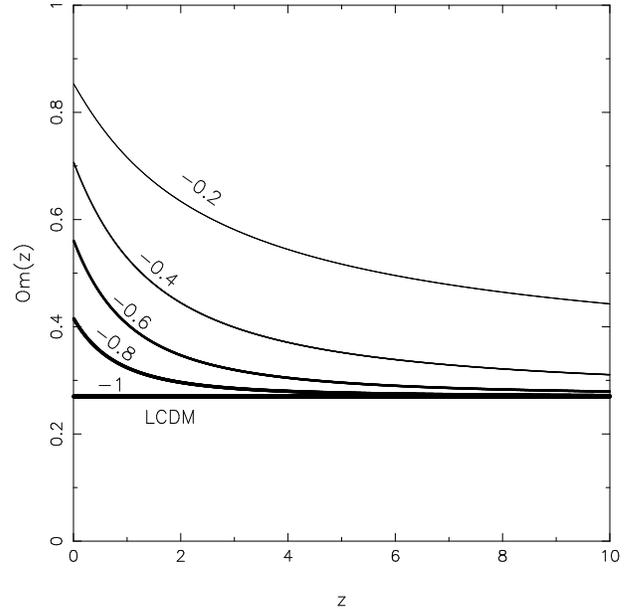}
\vspace{-0.6cm}
\end{center}
\caption{\small
The $Om$ diagnostic is shown as a function of redshift
for DE models with $\om=0.27$ and $w= -1, -0.8, -0.6, -0.4, -0.2$ (bottom to top).
For Phantom models (not shown) $Om$ would have the opposite curvature.}
\label{fig:om}
\end{figure}

\beq
Om(z) \equiv \frac{h^2(z)-1}{(1+z)^3-1}~,
\label{eq:om}
\eeq
where
\ber
h^2 = \frac{H^2(z)}{H_0^2} &=& \om (1+z)^3 + \Omega_{\rm DE}~,\nonumber\\
\Omega_{\rm DE} &=& (1-\om) \exp{\left\lbrace 3 \int_0^{z}
\frac{1+w(z')}{1+z'} dz'\right\rbrace }~.
\label{eq:hubble_recon}
\eer
is the expansion history of a spatially flat Friedmann-Robertson-Walker (FRW) Universe with
scale factor $a(t)$ and Hubble parameter $H(z)\equiv \dot a/a$.
The CPL ansatz expresses the EOS in terms of
the redshift $z$
in the following form \cite{CPL}:
\beq
w(z) = w_0 + w_1\frac{z}{1+z}.
\label{eq:cpl}
\eeq
In contrast to $w(z)$ and the deceleration parameter $q(z) \equiv
-{\ddot a}/aH^2$, the $Om(z)$ diagnostic depends upon no higher
derivative of the luminosity distance than the first one.
Therefore, it is less sensitive to observational errors than
either $w$ or $q$. $Om$ is also distinguished by the fact that
$Om(z) = \om$ for \lcdm.
$Om$ is very useful in establishing the properties of DE. For an
unevolving EOS: $1+w \simeq [Om(z) - \om](1-\om)^{-1}~$ at $z \ll
1$, consequently a larger $Om(z)$ is indicative of a larger $w$;
while at high $z$, $Om(z) \to \om$, as shown in figure
\ref{fig:om}.

\begin{figure*}[!t]
\centering
\begin{center}
\vspace{-0.05in}
\centerline{\mbox{\hspace{0.in} \hspace{2.1in}  \hspace{2.1in} }}
$\begin{array}{@{\hspace{-0.3in}}c@{\hspace{0.3in}}c@{\hspace{0.3in}}c}
\multicolumn{1}{l}{\mbox{}} &
\multicolumn{1}{l}{\mbox{}} \\ [-0.5cm]
 \includegraphics[scale=0.36, angle=-90]{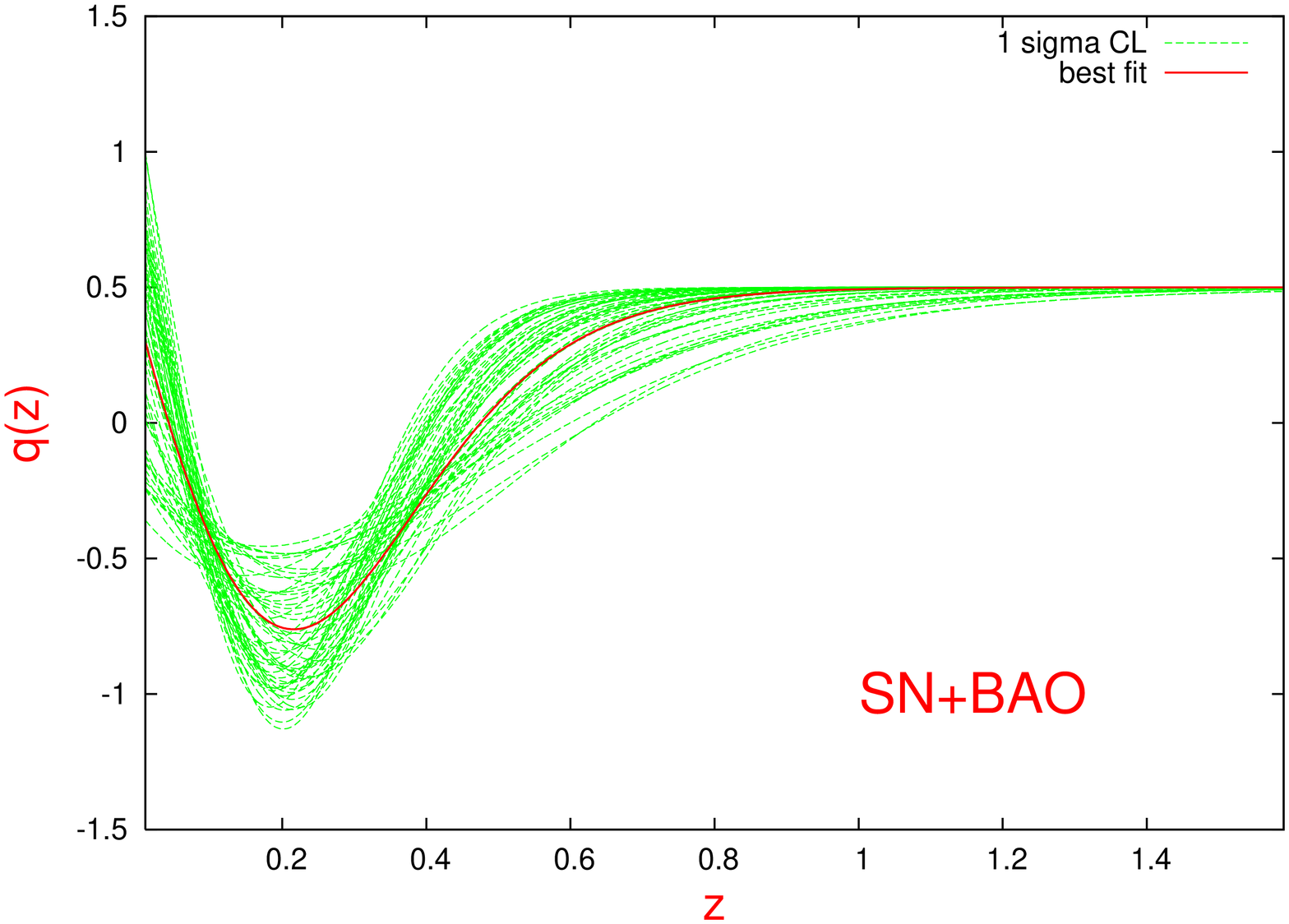}
\includegraphics[scale=0.36, angle=-90]{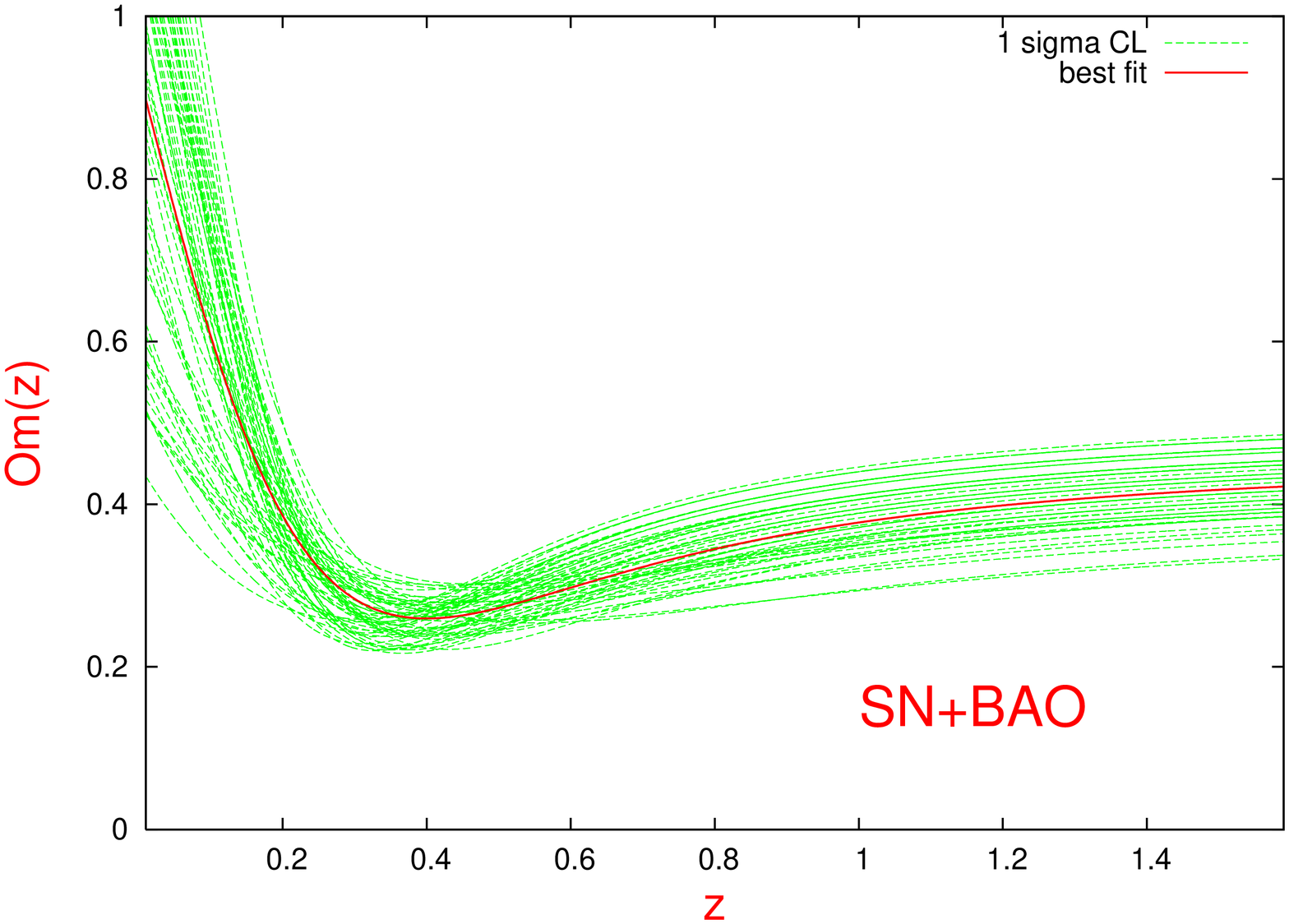}
\end{array}$
$\begin{array}{@{\hspace{-0.3in}}c@{\hspace{0.3in}}c@{\hspace{0.3in}}c}
\multicolumn{1}{l}{\mbox{}} &
\multicolumn{1}{l}{\mbox{}} \\ [-0.5cm]
 \includegraphics[scale=0.36, angle=-90]{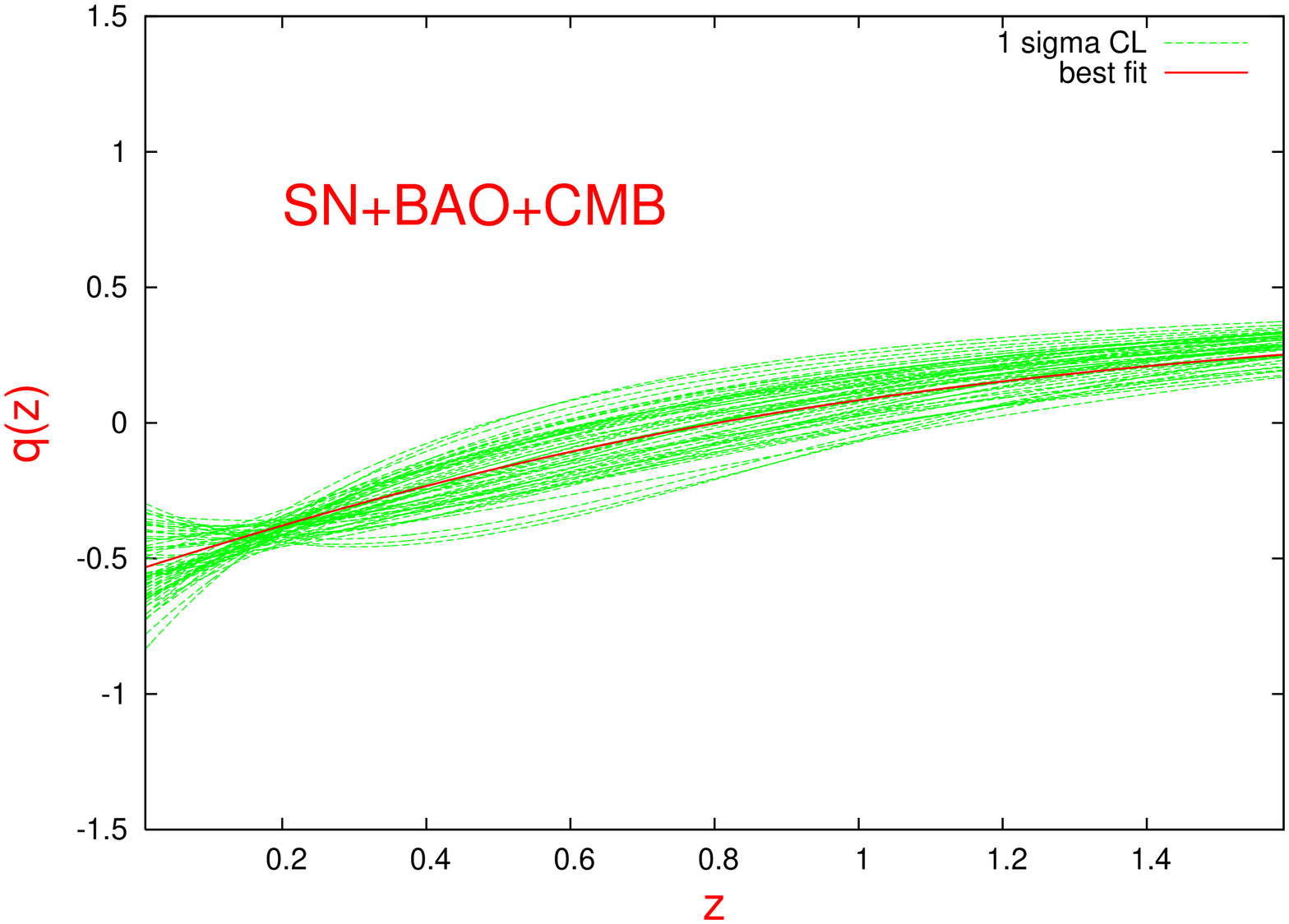}
\includegraphics[scale=0.36, angle=-90]{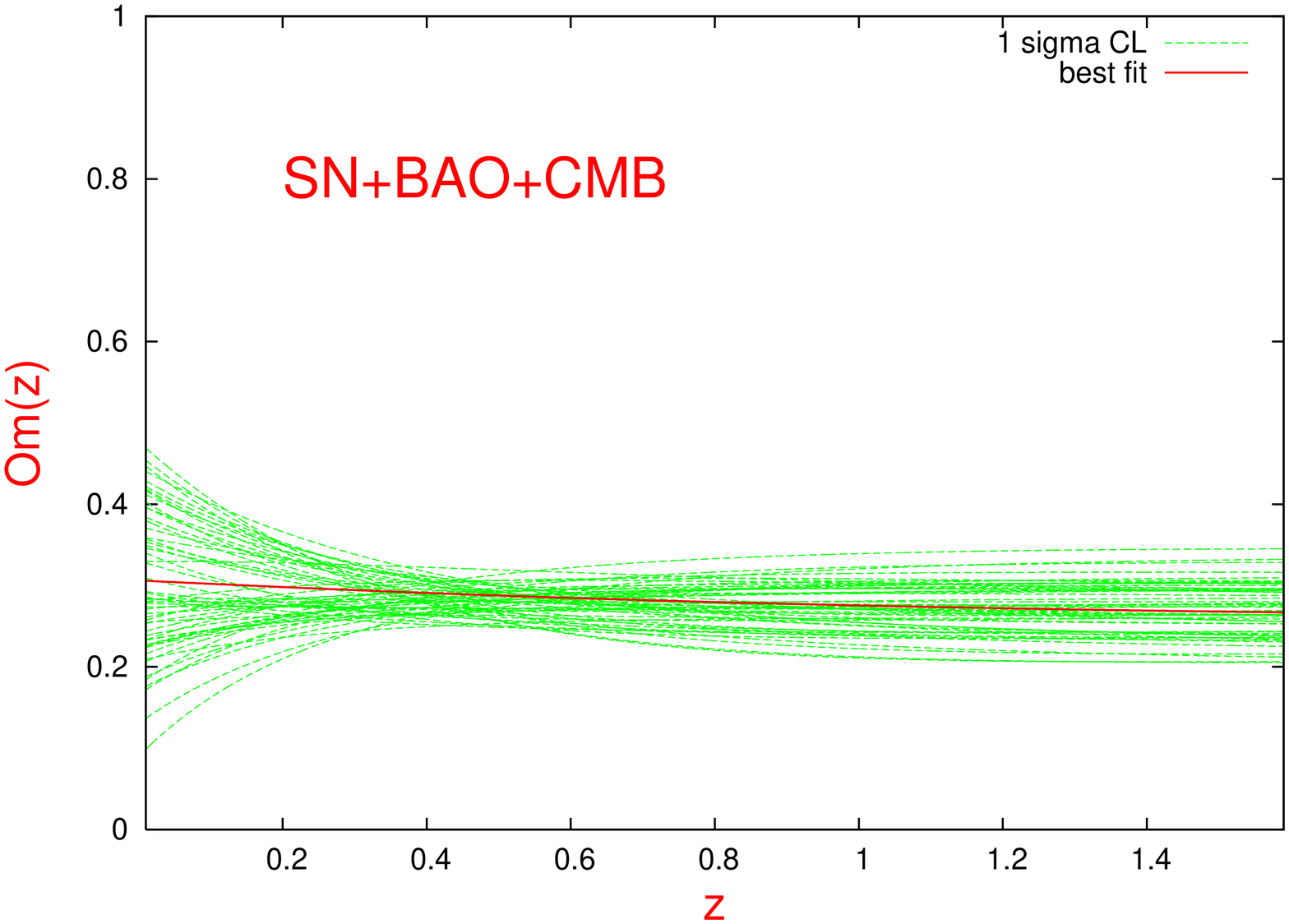}
\end{array}$
\end{center}
\caption{\small Reconstructed $q(z)$ and $Om(z)$
from SN Ia + BAO data (upper panels) and SN Ia + BAO + CMB data (lower panels) using
the CPL ansatz. Solid red lines show the best fit values of $Om(z)$ and $q(z)$
while dashed green lines show the $1\sigma$ CL.
The dramatic difference between the upper panels and the lower
one is indicative of the inability of the CPL parametrization to fit
the data at low and high redshifts simultaneously.
The spatially flat \lcdm model corresponds to a horizontal line with
$Om(z) = \om$ in the right panels
(not shown).} \label{fig_main}
\end{figure*}

The present analysis uses the recently compiled `Constitution set'
\cite{Constitution09}
of 397 type Ia supernovae covering a redshift range from
$z_{min}=0.015$ to $z_{max}=1.551$. The Constitution set is the largest SN Ia luminosity
distance sample currently available and includes 139 SN Ia at $z < 0.08$.
Our analysis considers the SN Ia data
individually as well as in combination with BAO distance
measurements obtained at $z=0.2$ and $z=0.35$ from the
joint analysis of the 2dFGRS and SDSS data \cite{BAO09}. The BAO distance ratio
$D_V(z=0.35)/D_V(z=0.20) = 1.736 \pm 0.065$ was shown in \cite{BAO09} to be
a relatively
model independent quantity. Here $D_V(z)$ is defined as
\beq
D_V(z_{BAO})= \bigg\lbrack~\frac{z_{BAO}}{H(z_{BAO})}~\bigg( \int_0^{z_{BAO}}
\frac{dz}{H(z)}~\bigg)^2~\bigg\rbrack^{1/3} .
\eeq
We also use the CMB shift parameter \cite{WMAP5} which is
the reduced distance to the last scattering surface ($z_{\rm ls} = 1090$)
\beq
R = \sqrt{\om}\int_0^{z_{\rm ls}} \frac{dz}{h(z)} = 1.71 \pm 0.019~.
\eeq

While SN Ia and BAO data contain information about the
Universe at relatively low redshifts, the $R$ parameter probes
the entire expansion history up to last scattering at
$z_{\rm ls}$. Because of this, our analysis also examines the
 goodness of fit for CPL parametrization when applied simultaneously to data at low and high redshifts.

Fig.~\ref{fig_main} shows
$q(z)$ and $Om(z)$ reconstructed using  (\ref{eq:cpl})
and SN Ia + BAO data (upper panels) and
SN Ia + BAO + CMB data (lower panels).
Red lines are best fit reconstructions
while dashed green lines show $1\sigma$ CL's from a $\chi^2$ analysis.
It is interesting that the best fit
flat $\Lambda$CDM model ($\Omega_{0m}=0.287$) satisfying SN Ia+BAO data
does not lie within the $1\sigma$ CL of our best reconstruction. By contrast, \cite{Constitution09}
obtain $1+w = 0.013^{+0.066}_{-0.068}$ after
assuming $w$ = constant, which underscores the difference made by dropping the
$w$ = constant constraint.
It is interesting to mention that the reduced $\chi^2$ also drops
from $\chi^2_{red}=1.182$ in case of $\Lambda$CDM model to $\chi^2_{red}=1.171$ in case of varying
dark energy model which makes the assumption of the additional parameter
worthwhile.

The growth in the value of $Om(z)$ at low $z$
 in the upper panel of
fig.~\ref{fig_main} is striking,
 and appears to favor a DE model with an EOS which increases
at late times. This could be preliminary evidence for a decaying DE model
since, in this case, the EOS would increase
at late times, resulting in an increase in the low $z$ value of the Hubble parameter
and therefore also of $Om(z)$.

\begin{figure*}[!t]
\centering
\begin{center}
\vspace{-0.05in}
\centerline{\mbox{\hspace{0.in} \hspace{2.1in}  \hspace{2.1in} }}
$\begin{array}{@{\hspace{-0.3in}}c@{\hspace{0.3in}}c@{\hspace{0.3in}}c}
\multicolumn{1}{l}{\mbox{}} &
\multicolumn{1}{l}{\mbox{}} \\ [-0.5cm]
 \includegraphics[scale=0.36, angle=-90]{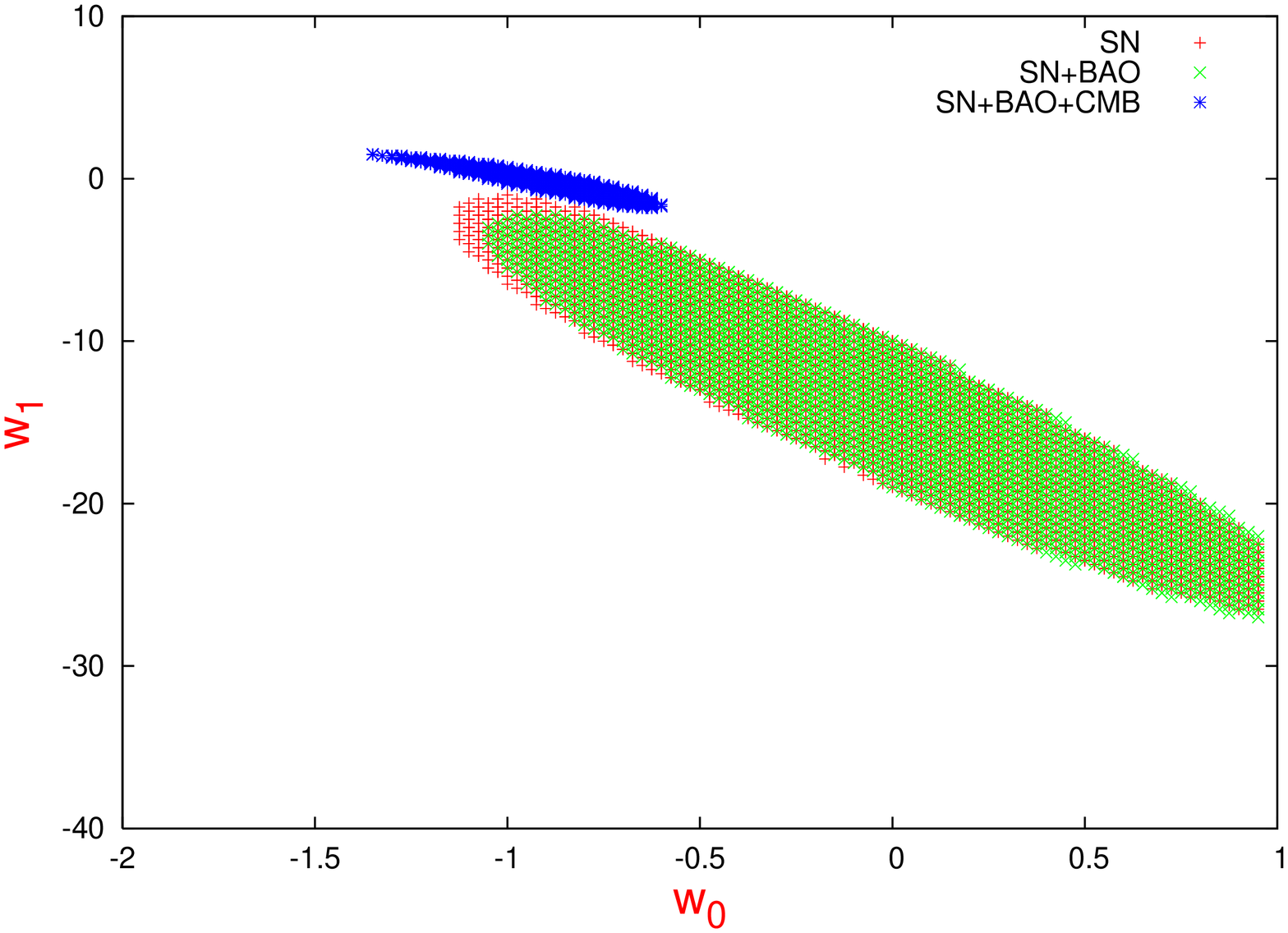}
 \includegraphics[scale=0.36, angle=-90]{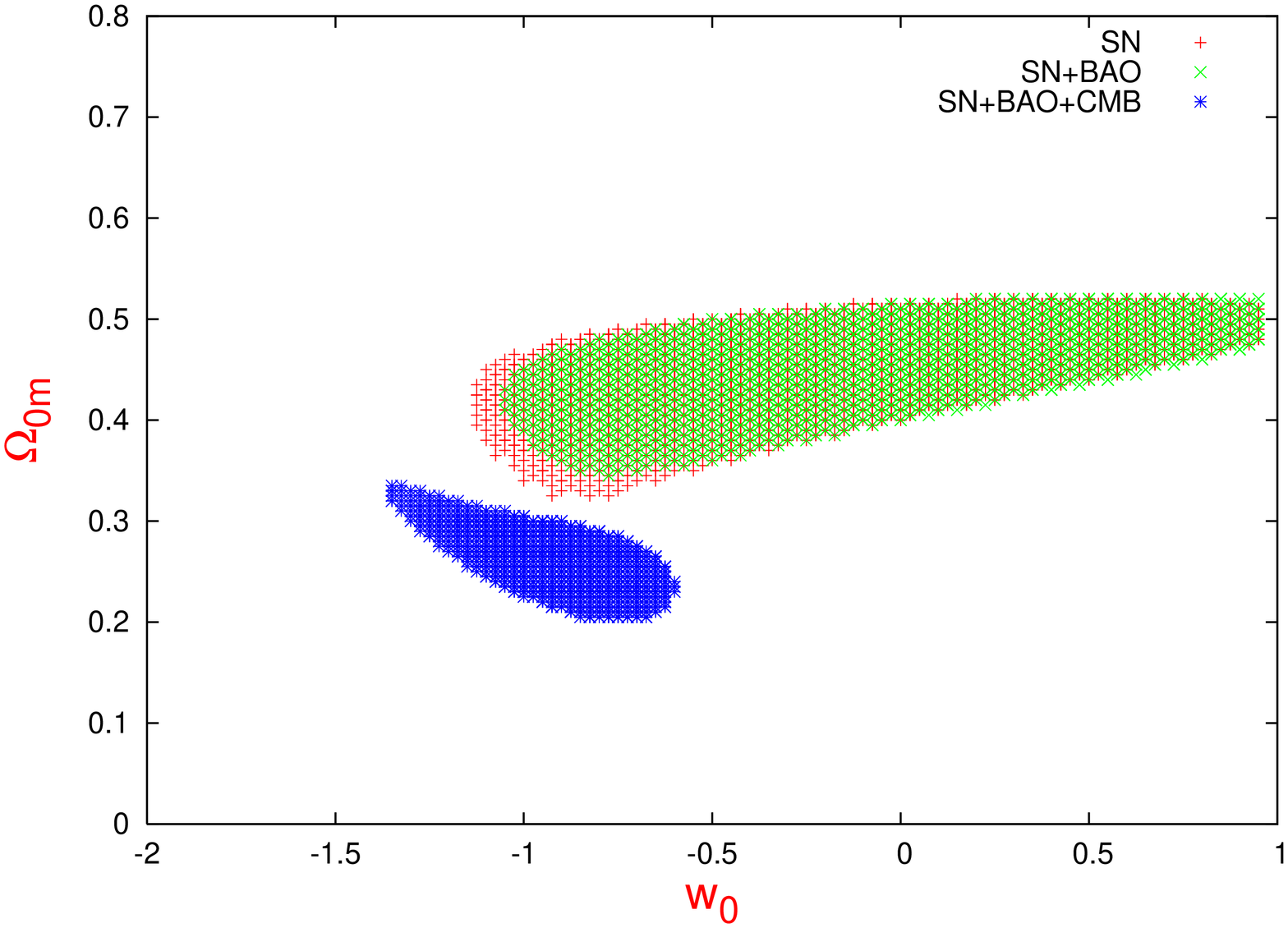}
\end{array}$
\end{center}
\caption{\small $1\sigma$ contours for CPL parameters $w_0$-$w_1$
(left panel) and $w_0$-$\Omega_{0m}$ (right panel) reconstructed
using SN Ia data (red pluses), SN Ia+BAO data (green crosses) and
SN Ia+BAO+CMB data (blue stars). Note the consistency between SN
Ia and BAO data. The absence of any overlap between the $1\sigma$
contours for SN Ia+BAO+CMB and SN Ia+BAO data could be indicative
of tension between the CPL parametrization and the data. We should
note here that plotting small contours that indicate tight
constraints on the parameters is misleading when the best fit
result produces a bad fit to the data (blue contours).}
\label{fig_contour}
\end{figure*}

These results change dramatically with the inclusion of CMB data.
The lower panel of fig.~\ref{fig_main} shows that our
reconstruction of $Om(z)$ is now perfectly consistent with
$\Lambda$CDM  -- for which $Om(z)$ is unevolving. However, one
should not stop and conclude that the standard model is confirmed
once more. Rather, what one observes here is an incompatibility of
the CPL parametrization for $w(z)$ with this combination of data
sets. In other words, the functional form of the CPL ansatz is
unable to fit the data simultaneously at low and high redshifts.
This can be clearly seen if we compare $\chi^2_{SN+BAO}=461.63$
{\bf ($\chi^2_{red}=1.171$)} for the best fit obtained using
SN+BAO data, with the significantly larger
$\chi^2_{SN+BAO+CMB}=467.07$ {\bf ($\chi^2_{red}=1.182$)} obtained
using SN+BAO+CMB(R) data. The  addition of one more data point
(the CMB shift parameter), increases the best fit $\chi^2$
 by more than $5$.
Support for this viewpoint is also provided by fig.~\ref{fig_contour}, which
shows the best fit regions in parameter space obtained using
the CPL ansatz after fitting to SN Ia (red pluses), SN Ia+BAO (green crosses) and
SN Ia+BAO+CMB (blue stars) data. Contrast the good overlap between
best fit regions obtained using SN Ia and SN Ia+BAO data,
with the relative isolation of the best fit region obtained using
SN Ia+BAO+CMB data.

Our reconstruction of $Om(z)$ appears to favor DE
with an increasing EOS 
at low redshifts $z\lleq 0.3$ (fig. \ref{fig_main} upper-right panel). We have also seen that the CPL
ansatz is strained to describe the DE behavior suggested by data at
low and high $z$. We believe the reason for
this stems from the fact that the CPL ansatz implicitly assumes that
the redshift interval from
$z=0$ to $z\sim 2$ represents nothing special for DE, so that $w(z)$
can safely be expanded in a Taylor series in powers of $z/(1+z)$ in this
interval.
This need not, however, be true, since models have been suggested in which
dark energy decays with a characteristic time of
order of the present age of the Universe \cite{DE_review,Decay,DS08}.
Note also that the large negative value of $w_1$ (fig.~\ref{fig_contour}) suggests that DE was
practically nonexistant at high redshifts. This too could be an ansatz-related feature,
since the reliability of (\ref{eq:cpl}) at high redshifts remains somewhat ambiguous.
Keeping these issues in mind, we re-analyze the data using a simple
toy model for $w(z)$ which encapsulates the main features of the effect
discovered, and show that this ansatz can provide a better fit to the
combination of SN+BAO+CMB data than CPL.

\begin{figure*}[!t]
\centering
\begin{center}
\vspace{-0.05in}
\centerline{\mbox{\hspace{0.in} \hspace{2.1in}  \hspace{2.1in} }}
$\begin{array}{@{\hspace{-0.3in}}c@{\hspace{0.3in}}c@{\hspace{0.3in}}c}
\multicolumn{1}{l}{\mbox{}} &
\multicolumn{1}{l}{\mbox{}} \\ [-0.5cm]
 \includegraphics[scale=0.36, angle=-90]{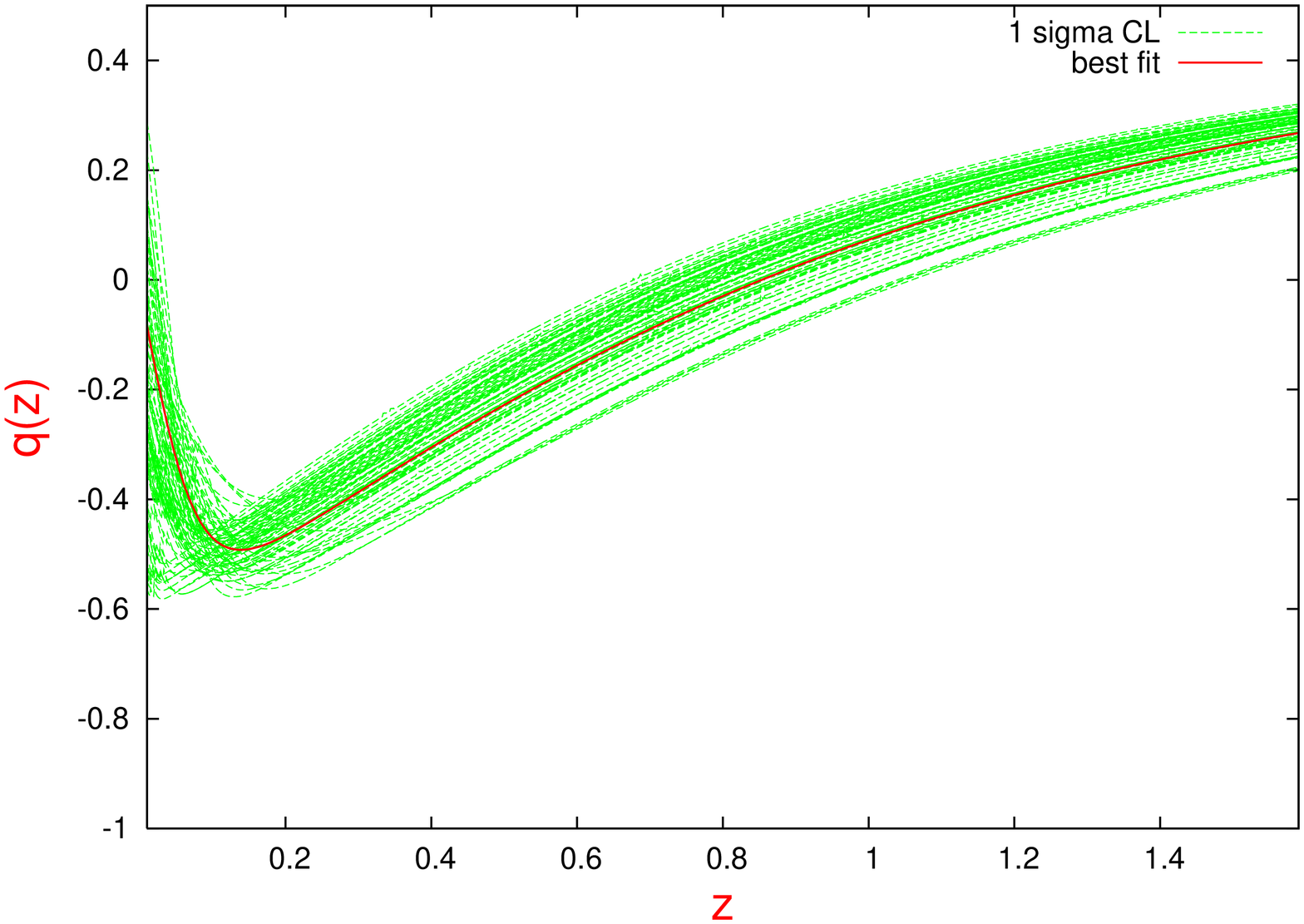}
 \includegraphics[scale=0.36, angle=-90]{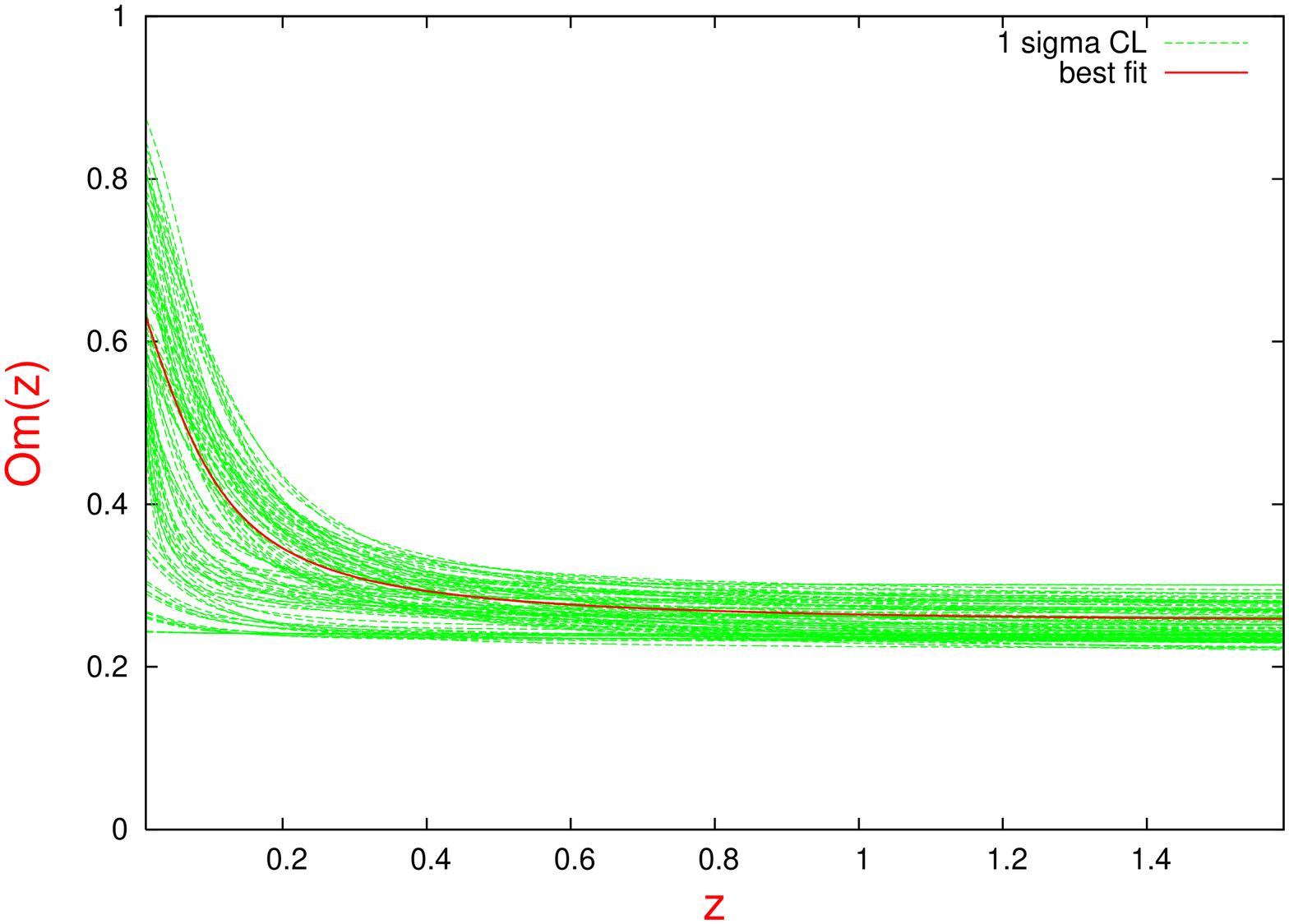}
\end{array}$
\end{center}
\caption{\small The cosmological deceleration parameter $q(z)$ (left panel)
and $Om(z)$ (right panel) reconstructed using a
combination of SN Ia, BAO and CMB data and the
ansatz (\ref{eq:step}).
Solid red lines show best fit reconstructed results
while dashed green lines show reconstructed results within $1\sigma$ CL.}
\label{fig_decay}
\end{figure*}

Our ansatz is
\beq
w(z)=- \frac{1+ \tanh\left[(z-z_t)\Delta\right]}{2}~,
\label{eq:step}
\eeq
(a similar form was used in \cite{bassett} and other papers
to search for fast phase transitions in DE at larger values of $z$). This fit
ensures $w = -1$ at early times, and then increases the EOS
to a maximum of $w\sim 0$ at low $z$.
It has the same number of
free parameters as the CPL ansatz but does not permit the crossing of the
phantom divide at $w = -1$.
The best fit cosmology obtained using this
ansatz has $z_t = 0.008$ (\ie $z_t\approx 0$), $\Delta = 12.8$, $\om = 0.255$ and
$\chi^2_{SN+BAO+CMB}=466.50$, and presents
an improvement ($\Delta \chi^2 = -0.6$)
over the best fit for the same data set obtained
using the CPL ansatz. 
We would like to emphasize
that this new fit is only used as an example to
demonstrate that the CPL ansatz may not be flexible enough to determine
cosmological parameters for a class of rapidly evolving DE models.
Fig.~\ref{fig_decay} shows the deceleration parameter
$q$ and the
$Om$ diagnostic reconstructed using (\ref{eq:hubble_recon}) \&
(\ref{eq:step}). Interestingly, \lcdm as well as a universe which is currently
coasting ($q_0 \simeq 0$), can both be accommodated by the data at roughly the same
level of confidence !

To check the robustness of our results we redo our analysis on
a subsample of the Constitution data set, namely the SNLS+ESSENCE+CfA
SN Ia data (234 data points in all).
Our results, shown in figure \ref{fig_subset}
for the CPL ansatz,
are summarized below:

\begin{figure*}[!t]
\centering
\begin{center}
\vspace{-0.05in}
\centerline{\mbox{\hspace{0.in} \hspace{2.1in}  \hspace{2.1in} }}
$\begin{array}{@{\hspace{-0.3in}}c@{\hspace{0.3in}}c@{\hspace{0.3in}}c}
\multicolumn{1}{l}{\mbox{}} &
\multicolumn{1}{l}{\mbox{}} \\ [-0.5cm]
\includegraphics[scale=0.36, angle=-90]{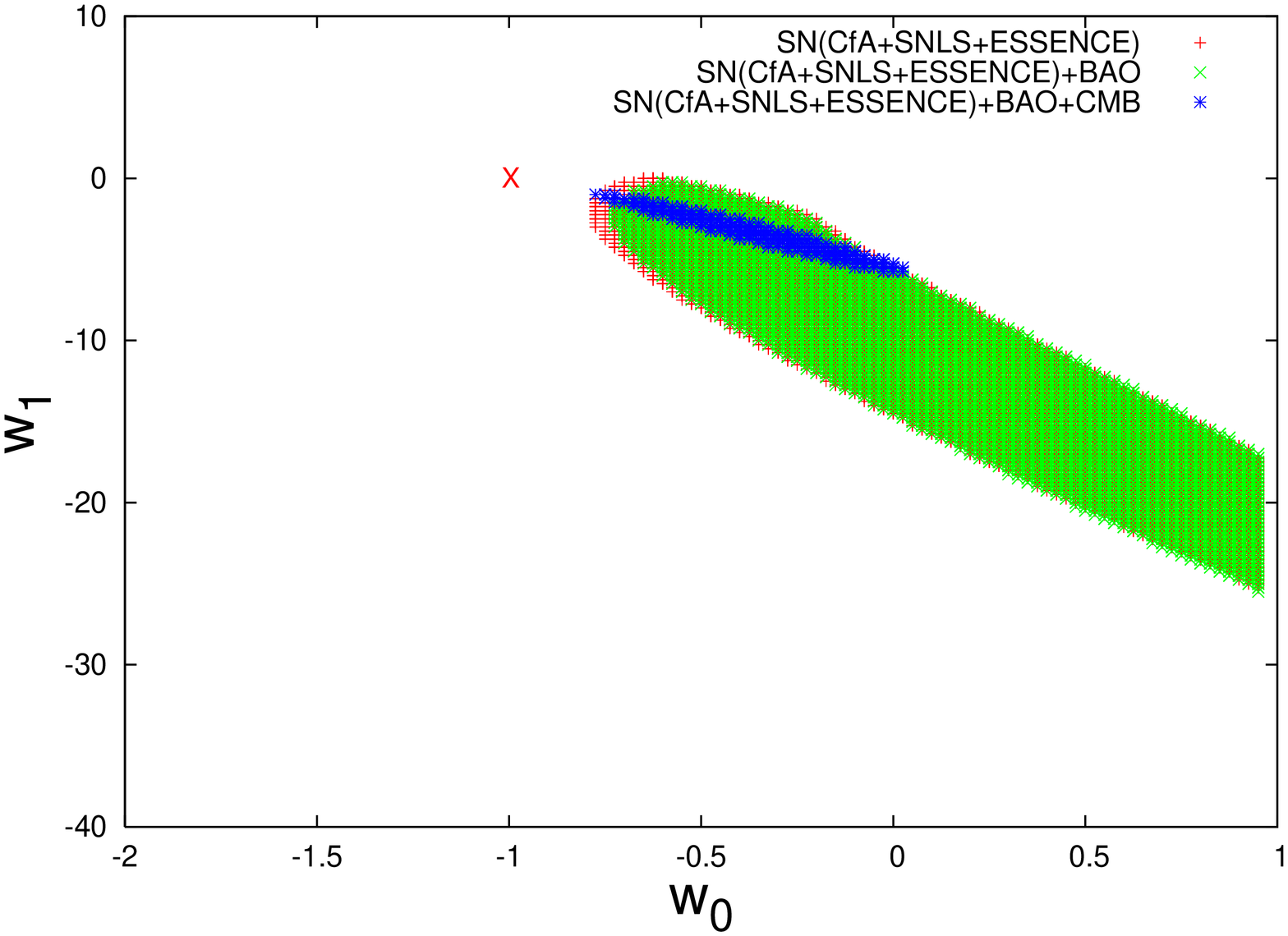}
\includegraphics[scale=0.36, angle=-90]{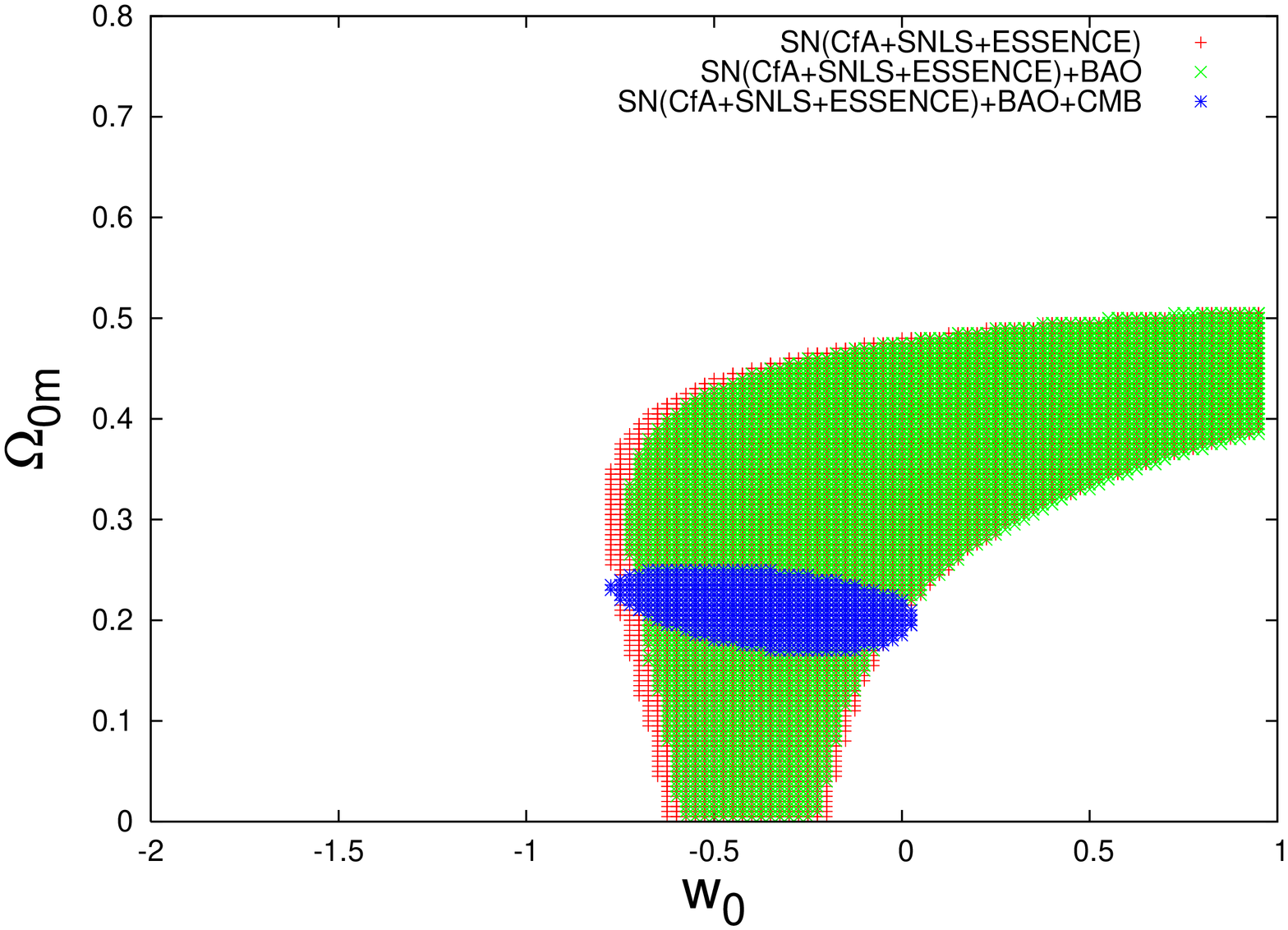}
\end{array}$
$\begin{array}{@{\hspace{-0.3in}}c@{\hspace{0.3in}}c@{\hspace{0.3in}}c}
\multicolumn{1}{l}{\mbox{}} &
\multicolumn{1}{l}{\mbox{}} \\ [-0.5cm]
 \includegraphics[scale=0.36, angle=-90]{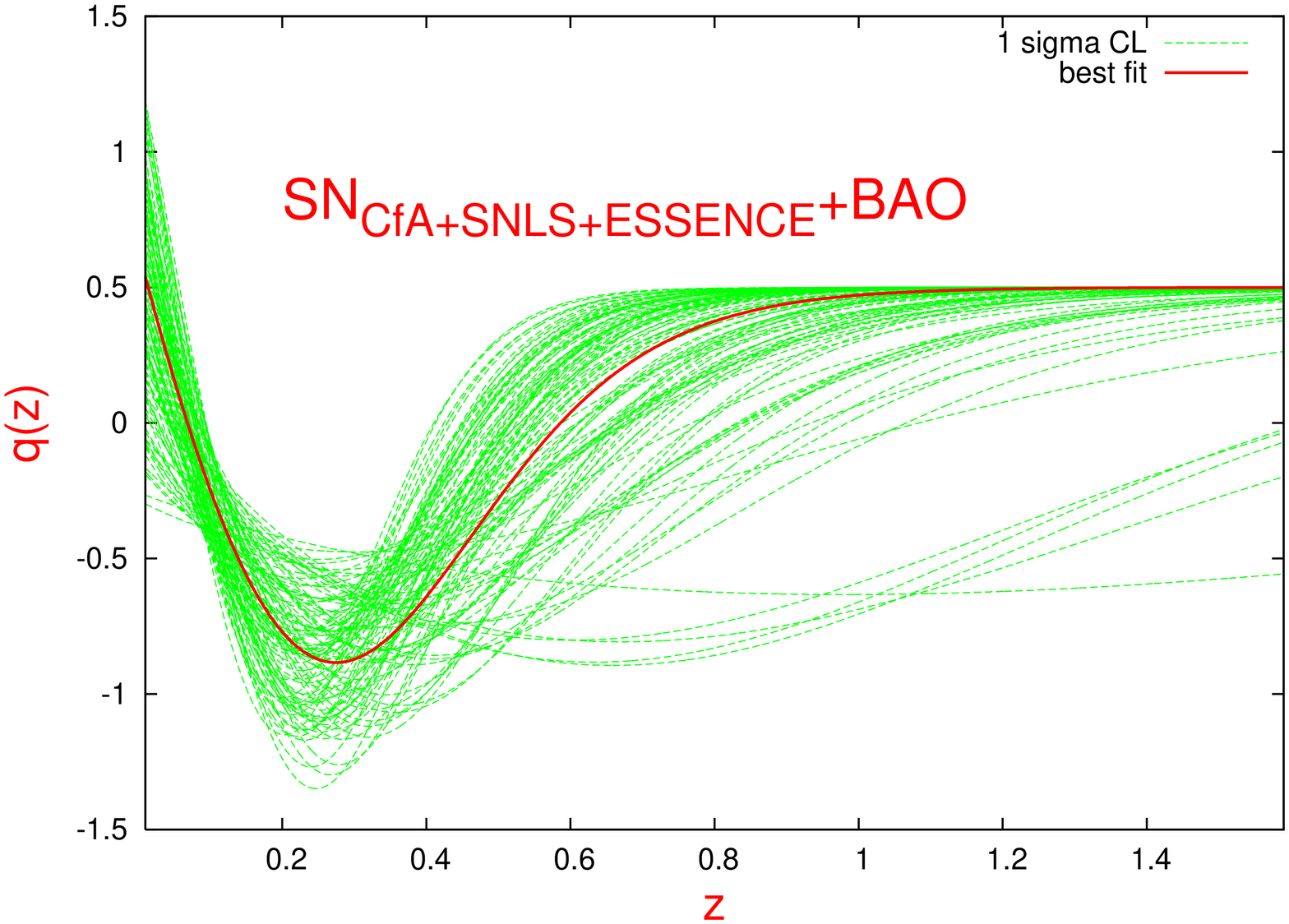}
\includegraphics[scale=0.36, angle=-90]{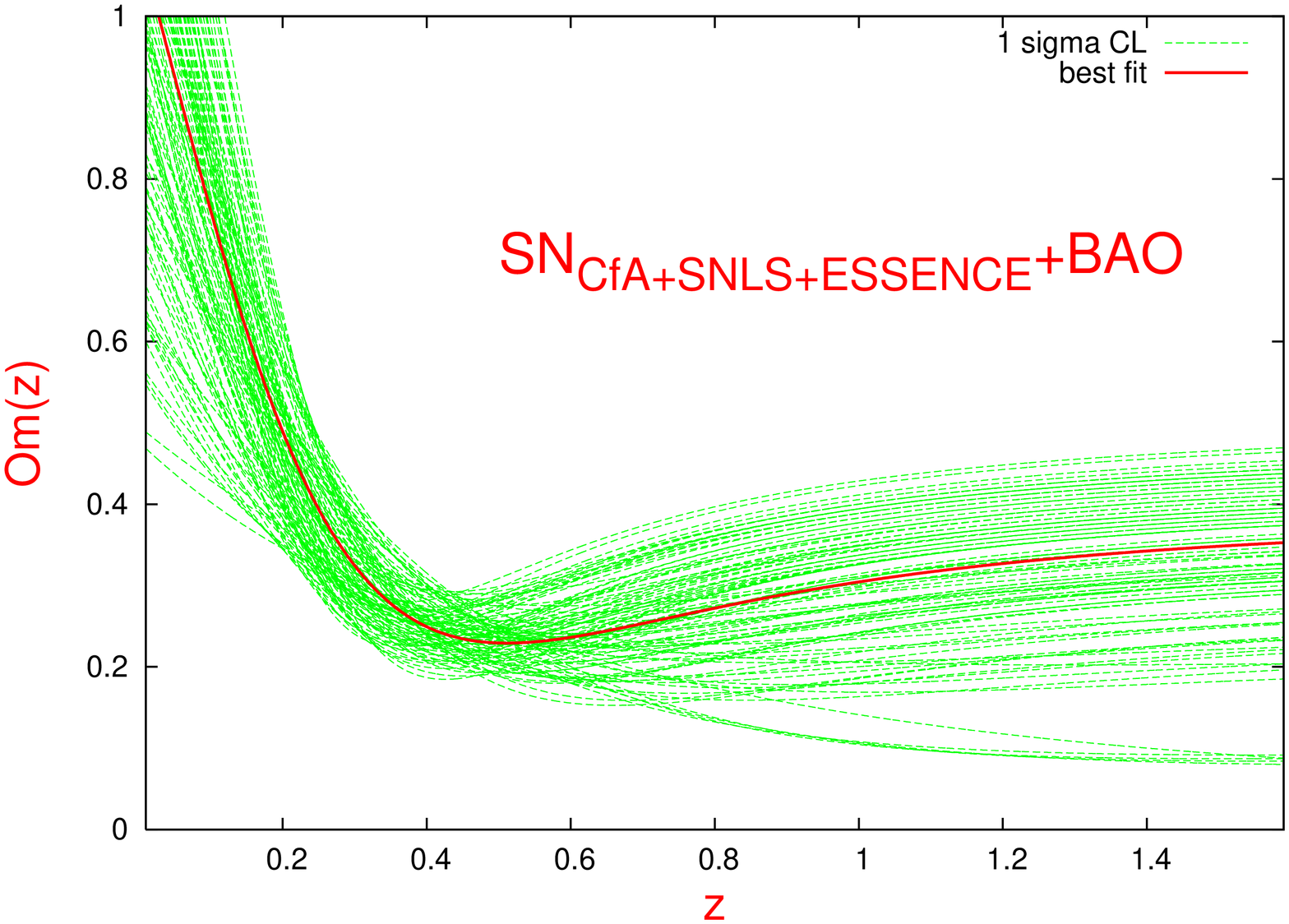}
\end{array}$
$\begin{array}{@{\hspace{-0.3in}}c@{\hspace{0.3in}}c@{\hspace{0.3in}}c}
\multicolumn{1}{l}{\mbox{}} &
\multicolumn{1}{l}{\mbox{}} \\ [-0.5cm]
 \includegraphics[scale=0.36, angle=-90]{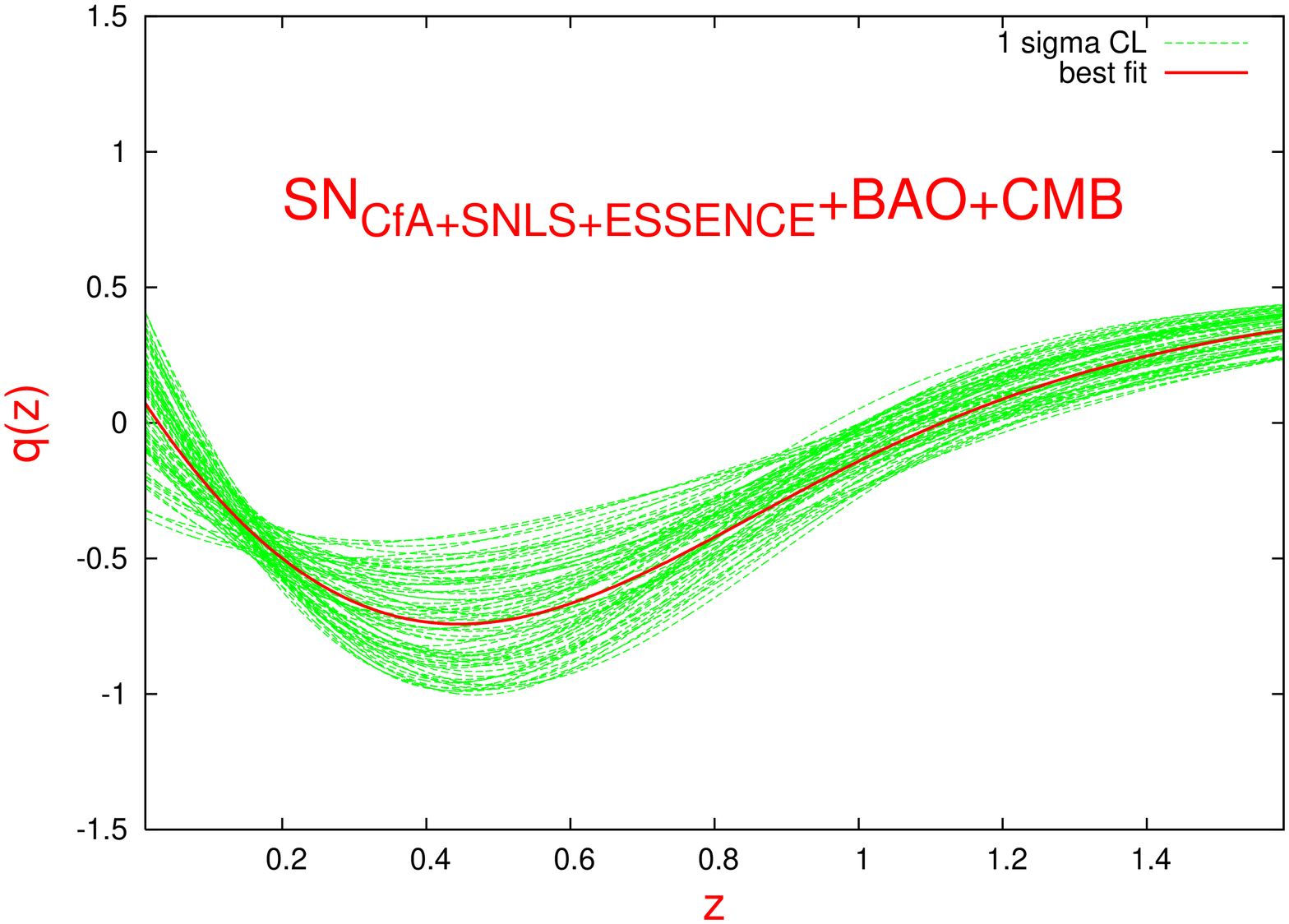}
\includegraphics[scale=0.36, angle=-90]{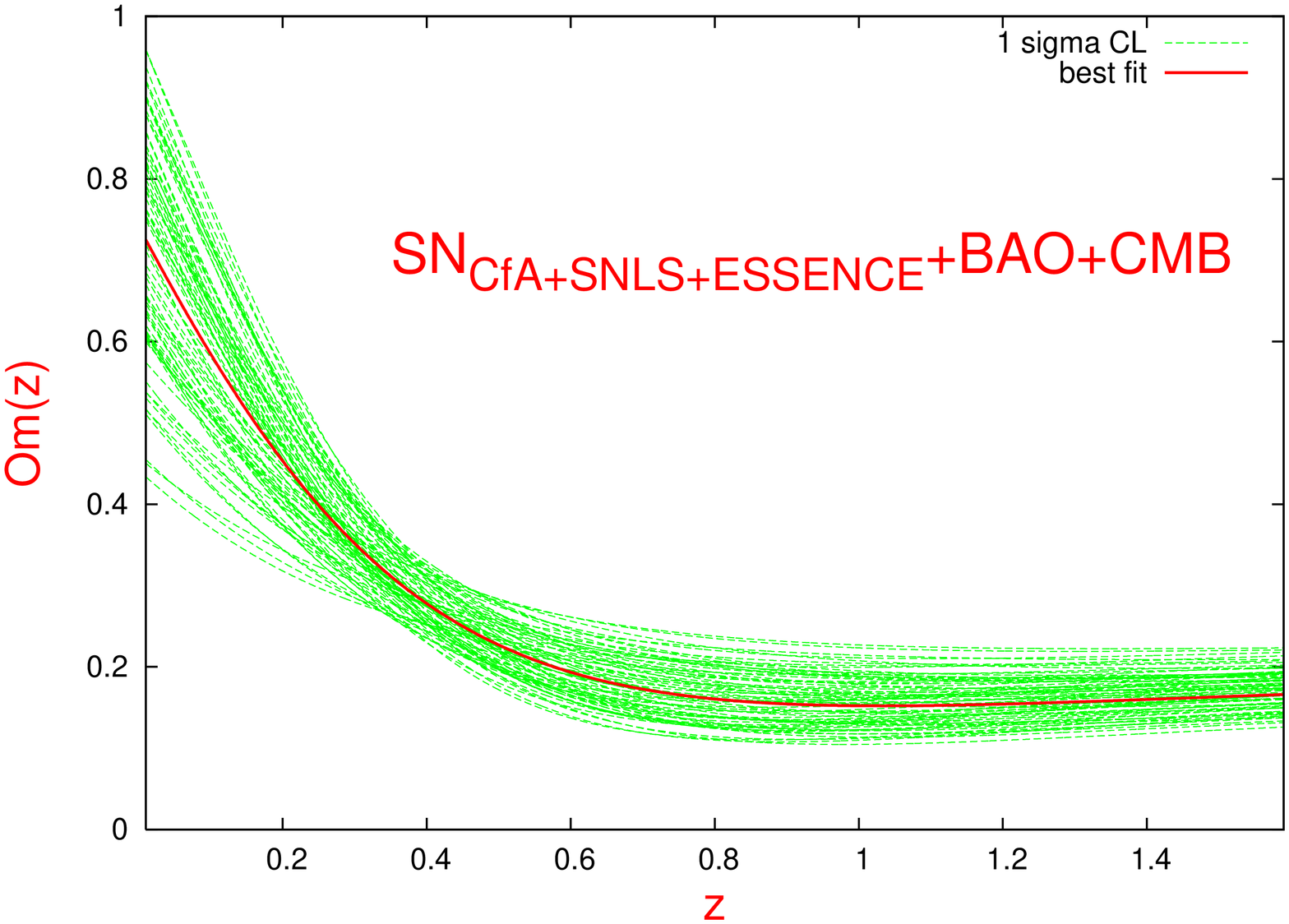}
\end{array}$
\end{center}
\caption{\small {\bf Top:}
$1\sigma$ contours for CPL parameters $w_0$-$w_1$ (left panel) and
$w_0$-$\Omega_{0m}$ (right panel) reconstructed using SN Ia data
(SNLS+ESSENCE+CfA, red pluses), SN Ia+BAO data (green crosses) and SN Ia+BAO+CMB data (blue stars). Note the compatibility between the different data sets and that spatially flat \lcdm (red cross at $w_0=-1, w_1=0$ in the left panel) appears to be in tension with this combination of data.
{\bf Middle and Bottom:} Reconstructed $q(z)$ and $Om(z)$
from SN Ia + BAO data (middle). Reconstructed $q(z)$ and $Om(z)$ from SN Ia + BAO + CMB data (bottom). In the middle and bottom  panels solid red lines show the best fit values of $Om(z)$ and $q(z)$ while dashed green lines show the $1\sigma$ CL. In the bottom panel, the best fit values for \lcdm are: $q_0 \simeq -0.6$ and $Om(z) = 0.255$. All results have been obtained using the CPL ansatz (\ref{eq:cpl}).}
\label{fig_subset}
\end{figure*}

\begin{enumerate}

\item  An excellent overlap exists between the 1$\sigma$ contours
$w_0$-$w_1$ (top left) and $w_0$-$\Omega_{0m}$ (top right)
reconstructed using SN Ia, SN Ia+BAO and SN Ia+BAO+CMB data. This
demonstrates that the CPL ansatz works quite well for this
combination of data sets and hints that the tension noticed in
figure \ref{fig_contour} could be coming from data sets which have
been excluded from the present SN Ia compilation: namely the Gold
data, the high $z$ HST data and older SN Ia data sets. The visual
impression conveyed by this panel, which appears to support
evolving DE, receives statistical support: the best fit $\chi^2$
for SN Ia data is 267.69, while $\chi^2 = 267.92$ for SN Ia+BAO
and $\chi^2 = 268.89$ for SN Ia+BAO+CMB. We therefore find that
the $\chi^2$ values for the three data sets (SN Ia, SN Ia+BAO, SN
Ia+BAO+CMB) lie much closer together for the data shown in figure
\ref{fig_subset}, compared to the data in figure
\ref{fig_contour}.

\item A larger value of $Om(z)$ and $q(z)$ at low redshifts is supported by
the present analysis of
SNLS+ESSENCE+CfA supernovae in combination with BAO and CMB data (figure \ref{fig_subset}
middle and bottom).
Indeed, coasting cosmology ($q_0 \simeq 0$)
provides an excellent fit to the data while $\Lambda$CDM ($q_0 \simeq -0.6)$ appears to be excluded
at 1$\sigma$ by this data set.
This is in marked contrast to the results in fig. \ref{fig_main} (lower panel)
which distinctly favoured \lcdm.

\item The fact that the spatially flat \lcdm shows weaker
consistency with the SN Ia subsample + BAO + CMB data is clearly
seen from the best fit values for \lcdm: (i) $\chi^2 = 274.64$
using only SN Ia ($\Omega_m = 0.28$), (ii) $\chi^2 = 275.87$ for
SN Ia+BAO($\Omega_m = 0.28$), (iii) $\chi^2 = 276.84$ for SN
Ia+BAO+CMB ($\Omega_m = 0.26$). Comparing with the results for
evolving DE discussed earlier, we find that the incremental value
of $\Delta \chi^2$ between the best fit evolving DE model and best
fit $\Lambda$CDM is $\simeq 8$, and favours evolving DE (the
reduced $\chi^2$ drops from $\chi^2_{red}=1.188$ in case of
$\Lambda$CDM model to $\chi^2_{red}=1.159$ in case of varying dark
energy model).

\end{enumerate}

To summarize, the recently released
Constitution SN Ia data set appears to support DE evolution at low redshifts.
There also appears to be some tension between low $z$
(Constitution SN Ia + BAO) and high $z$ (CMB) data, when analyzed
using the CPL ansatz. (However, this tension decreases when only a
subsample of the Constitution set is analysed.) There could be
several reasons for this.

\begin{itemize}

\item Systematics in some of the data sets is not sufficiently
well understood. This effect may have a purely astronomical
explanation and be a result of some systematic effect, e.g. if new
nearby CfA SN Ia are brighter on average. However, we have found
additionally found that if the effect is assumed to be
cosmological, then the implied DE behavior at low redshifts (using
SN Ia data) is more consistent with a rather large value of the
$D_V(z=0.35)/D_V(z=0.20)$ BAO distance ratio derived in
\cite{BAO09}.

\item  Another possibility is that this behavior of $Om,~w,~q$ is
an {\em apparent} one, which is induced by a local spatial
inhomogeneity -- a kind of a ``Hubble bubble", but with a
large-scale matter overdensity (though the results of
\cite{Constitution09} do not appear to support its existence).

\item Different SN Ia subsamples comprising the Constitution set
have varying properties.

\item The CPL ansatz is not versatile enough to accommodate the
cosmological evolution of dark energy suggested by the data.

\end{itemize}

Clearly one must wait for more data before deciding between these
alternatives. Note that the bulwark of support for evolving DE
comes from BAO data in conjunction with Constitution SN Ia data
\cite{Constitution09}  which includes 139 SN Ia at $z < 0.08$. It
is therefore encouraging that several hundred SN Ia light curves
are expected to be presented 
by the Nearby Supernova Factory ($0.03 < z < 0.08$) and SDSS ($0.05 < z < 0.35$).
As concerns $\Lambda$CDM, our conclusion is
that it is still viable 
at the $95\%$ confidence level but perhaps one should take its
subtle challenges more seriously \cite{BND}.

AS acknowledge BIPAC and the support of the EU FP6 Marie Curie Research and
Training Network ``UniverseNet" (MRTN-CT-2006-035863) and thanks
Pedro Ferreira, Mark Sullivan and David Polarski
for the useful discussions. AAS acknowledges RESCEU hospitality as a visiting
professor. He was also partially supported by the grant RFBR 08-02-00923 and
by the Scientific Programme ``Astronomy'' of the Russian Academy of Sciences.

\end{document}